**Defect-Engineered Layer-Dependent Nonlinear Optical Response in Two-Dimensional Muscovite for Efficient Optical Limiting**

*Dipanwita Mitra, Guilherme S. L. Fabris, Raphael Benjamim, Riya Sadhukhan, Jayanta Kumar Sarkar, Mateus M. Ferrer, Marcelo Lopes Pereira Junior, Preeti Lata Mahapatra, Dipak Kumar Goswami, Gelu Costin, Douglas S. Galvão*, Chandra Sekhar Tiwary*, Prasanta Kumar Datta**

Dipanwita Mitra, Riya Sadhukhan, Jayanta Kumar Sarkar, Dipak Kumar Goswami, Prasanta Kumar Datta
Department of Physics
Indian Institute of Technology Kharagpur
Kharagpur 721302, India
E-mail: pkdatta@phy.iitkgp.ac.in

Preeti Lata Mahapatra
School of Semiconductor Manufacturing
New Age Makers Institute of Technology
Gandhinagar, Gujarat, India

Guilherme S. L. Fabris, Douglas S. Galvao
Department of Applied Physics and Center for Computational Engineering and Sciences,
State University of Campinas
Campinas, São Paulo, Brazil
E-mail: galvao@ifi.unicamp.br

Raphael Benjamim, Marcelo Lopes Pereira
Department of Materials Sciences and Nano Engineering,
Rice University, Houston, Texas, USA

Mateus M. Ferrer
Graduate Program in Materials Science and Engineering, Technological Development Center
Federal University of Pelotas
Pelotas, Rio Grande do Sul, Brazil
E-mail:mateusmferrer@gmail.com

Gelu Costin
Department of Earth Environmental and Planetary Sciences
Rice University
Houston, Texas 77005, USA

Chandra Sekhar Tiwary
Department of Metallurgical and Materials Engineering
Indian Institute of Technology Kharagpur
Kharagpur, 721302, India
E-mail: chandra.tiwary@metal.iitkgp.ac.in

**Abstract text**

Light–matter interactions in two-dimensional (2D) materials have gained significant interest due to their distinctive optical and electronic properties. Recently, silicates have emerged as a promising new class of 2D materials, but their nonlinear optical properties remain largely unexplored. This study demonstrates the layer-dependent nonlinear absorption and optical limiting capabilities of 2D muscovite, a silicate mineral, using femtosecond laser excitation at 450 nm. The two-photon absorption (TPA) coefficient is highly sensitive to both number of layers and excitation intensity, increasing markedly from $(3.91 \pm 0.06) \times 10^3$ cm/GW in multilayer structures to $(6.94 \pm 0.17) \times 10^5$ cm/GW in the monolayer limit at a peak intensity 68 GW/cm$^2$, highlighting a strong layer-dependent enhancement in nonlinear absorption. Additionally, monolayer muscovite exhibits an optical limiting threshold of 1.46 mJ/cm$^2$, outperforming graphene and other 2D dichalcogenides. This enhanced TPA results from quantum confinement and intrinsic lattice defects that facilitate nonlinear optical transitions. Density functional theory reveals that liquid-phase exfoliation disrupts potassium interlayers and induces oxygen vacancies, creating mid-gap electronic states that significantly enhance TPA. These insights open new avenues for designing low-fluence, high-efficiency optical limiters using 2D silicates.

## 1. Introduction

Nonlinear optics is a branch of optics that explores the behavior of light in nonlinear media[1,2]. Nonlinear optical (NLO) materials are becoming increasingly important across the entire spectrum of photonic technologies, from light generation and control to transmission, detection, and imaging [1–4]. These materials serve as the backbone of numerous widely used photonic devices, such as pulsed lasers[5], optical switches[6], modulators[7], photodetectors[8], and optical memories[9] underscoring the distinct advantages of optical technologies over standard electronic systems.

Increasing concerns about laser-induced damage to human eyes and optical systems have accelerated the demand for advanced optical limiting (OL) materials. An effective optical limiter selectively transmits low-intensity light while blocking high-intensity light, making it of fundamental importance for applications such as passive mode locking[10], pulse shaping[11], and eye protection.

OL is driven by nonlinear effects, including nonlinear scattering (NLS), absorption (NLA), and refraction (NLR). Among absorption mechanisms, key processes include free-carrier absorption (FCA), reverse saturable absorption (RSA), and multiphoton absorption (MPA). Notably, two-photon absorption (TPA) enables materials to absorb photons with energies below the bandgap values. In two-dimensional (2D) materials, quantum confinement leads to bandgap opening as the particle size falls below the exciton Bohr radius, causing a blueshift in the absorption peaks [12]. This broadens the transparency window and strengthens nonlinear optical (NLO) responses, thereby enhancing optical limiting performance through the synergistic effects of multiphoton and FCA [12].

In recent years, the study of OL materials has increasingly turned toward 2D materials. Among them, graphene and its derivatives have shown exceptional potential, thanks to their linear electronic band structure and extended conjugated $sp^2$ π-system, which enable ultrafast carrier relaxation and a broadband NLO response[13–15]. These intrinsic characteristics make graphene a highly promising candidate for OL applications in both solution-based and solid-state systems. The OL performance of graphene nanostructures, such as graphene oxide nanosheets (GONSs) and graphene nanosheets (GNSs), has been systematically investigated using the open-aperture Z-scan technique with nanosecond laser pulses at 532 nm and 1064 nm [13]. At an input energy of 250 μJ/pulse, GONSs exhibit an OL threshold exceeding 3 J/cm² at 532 nm, while GNSs display significantly lower thresholds of 0.5 J/cm² at 532 nm and 6.3 J/cm² at 1064 nm [13]. Remarkably, single-layer graphene has achieved an OL threshold as low as 10 mJ/cm² at 532 nm[16].

Beyond graphene, a broad spectrum of 2D materials, such as black phosphorus (BP)[17] ,antimonene[18], hexagonal boron nitride (h-BN)[19], halide perovskites[20], transition metal dichalcogenides (TMDs)[21,22], metal oxides[23], layered double hydroxides (LDHs)[24], and metal-organic frameworks (MOFs)[25], have been extensively explored for their NLO and OL properties. $WS_2$ films with 1–3 layers have demonstrated a two-photon absorption (TPA) coefficient of $(1.0 \pm 0.8) \times 10^4$ cm/GW at 1030 nm under femtosecond excitation [26], while monolayer $MoS_2$ exhibited a TPA coefficient of $(7.62 \pm 0.15) \times 10^3$ cm/GW under similar conditions [21]. Notably, an annealed BP:$C_{60}$/PMMA composite film showed enhanced OL performance at 532 nm, attributed to thermally induced intermolecular charge transfer between BP and $C_{60}$, achieving a nonlinear absorption coefficient of 241.73 cm/GW, an OL threshold of 4.5 J/cm², and a damage threshold of 19.54 J/cm² [27].
2D silicates have rapidly emerged as outstanding candidates for NLO applications, showcasing remarkable potential not only in photonics and optoelectronics but also across a broad spectrum of cutting-edge technological fields [12,28,29]. The nonlinear optical phenomenon of second harmonic generation (SHG) was first demonstrated in single-crystal $SiO_2$ [30], marking a pivotal advancement in the field. Building on this foundation, first-principles predictions of SHG in non-centrosymmetric silicate crystals were successfully validated through experimental studies [31], further highlighting the strong potential of silicates for NLO applications.
In our previous study, we demonstrated the remarkable NLO performance of 2D biotite, a naturally occurring layered silicate. Monolayer biotite exhibited an outstanding TPA coefficient of $(9.75 \pm 0.15) \times 10^5$ cm/GW at 415 nm under femtosecond laser excitation at a peak intensity of 12 GW/cm²[12]. Furthermore, it exhibited a notably low OL threshold of 1.51 mJ/cm², surpassing several widely studied 2D materials, including graphene and TMDs. These findings underscore the exceptional potential of 2D silicates for next-generation photonic and optoelectronic technologies.
This study presents the synthesis of ultrathin 2D muscovite, another promising layered silicate mineral, via liquid-phase exfoliation of the bulk material, a phyllosilicate with the general formula $K\,AL_2\left(Si_3\,Al\right)O_{10}\left(OH\right)_2$.Exfoliation is performed over 2, 4, and 6 hours to examine the effect of sonication time systematically. The layer-dependent NLA and OL properties of 2D muscovite are characterized using the open-aperture (OA) Z-scan technique. Atomic force microscopy (AFM) is used to determine the thickness and lateral dimension of the nanosheets, while structural and optical characterizations are performed using UV–Vis spectroscopy, Raman spectroscopy, high-resolution transmission electron microscopy (HRTEM), scanning electron microscopy (SEM), X-ray diffraction (XRD) and X-ray photoelectron spectroscopy (XPS). Zeta potential measurements assess the surface charge of the exfoliated nanosheets. The TPA coefficient exhibits a strong dependence on both the number of layers and the excitation intensity, increasing significantly from $(3.91 \pm 0.06) \times 10^3$ cm/GW in the 2h exfoliated sample (12–13L) to $(6.94 \pm 0.17) \times 10^5$ cm/GW in the 6h exfoliated monolayer sample, at a common peak intensity of 68 GW/cm² and 450 nm excitation wavelength. This substantial increase highlights the pronounced layer-dependent enhancement in nonlinear optical absorption. To further elucidate the nonlinear optical behaviour and support the interpretation of the experimental data, density functional theory (DFT) simulations are also performed.

## 2. Results and Discussion

Muscovite is a phyllosilicate (sheet silicate) mineral characterized by a TOT-c structure. In simple terms, its crystal lattice consists of stacked TOT layers held together by interlayer potassium cations ($K^+$) [32]. **Figure S1a** illustrates the crystal structure of a muscovite unit cell. Each TOT layer comprises three sheets: two outer tetrahedral (T) sheets and one central octahedral (O) sheet. The T sheets consist of silicon-oxygen and aluminium-oxygen tetrahedra, where three of the four oxygen anions in each tetrahedron are shared with neighbouring tetrahedra, forming a hexagonal arrangement. The fourth oxygen, pointing inward, is known as the apical oxygen anion [32]. The central O sheet contains aluminium cations coordinated by six oxygen or hydroxide anions, forming octahedra. These octahedra also form a hexagonal layer by sharing anions. The apical oxygen anions from the adjacent T sheets are shared with the octahedral sheet, creating strong inter-sheet bonding within each TOT layer. This robust intra-layer bonding contrasts with the comparatively weaker inter-layer bonding provided by potassium cations, giving muscovite its characteristic perfect basal cleavage.

Muscovite, a naturally layered mineral, readily exfoliates along its basal planes (001). During exfoliation, the layer thickness gradually decreases, resulting in ultrathin materials that approach the 2D limit. This transformation is demonstrated in **Figure S1b**, which illustrates the liquid-phase exfoliation of bulk muscovite into 2D nanosheets using ultrasonication.

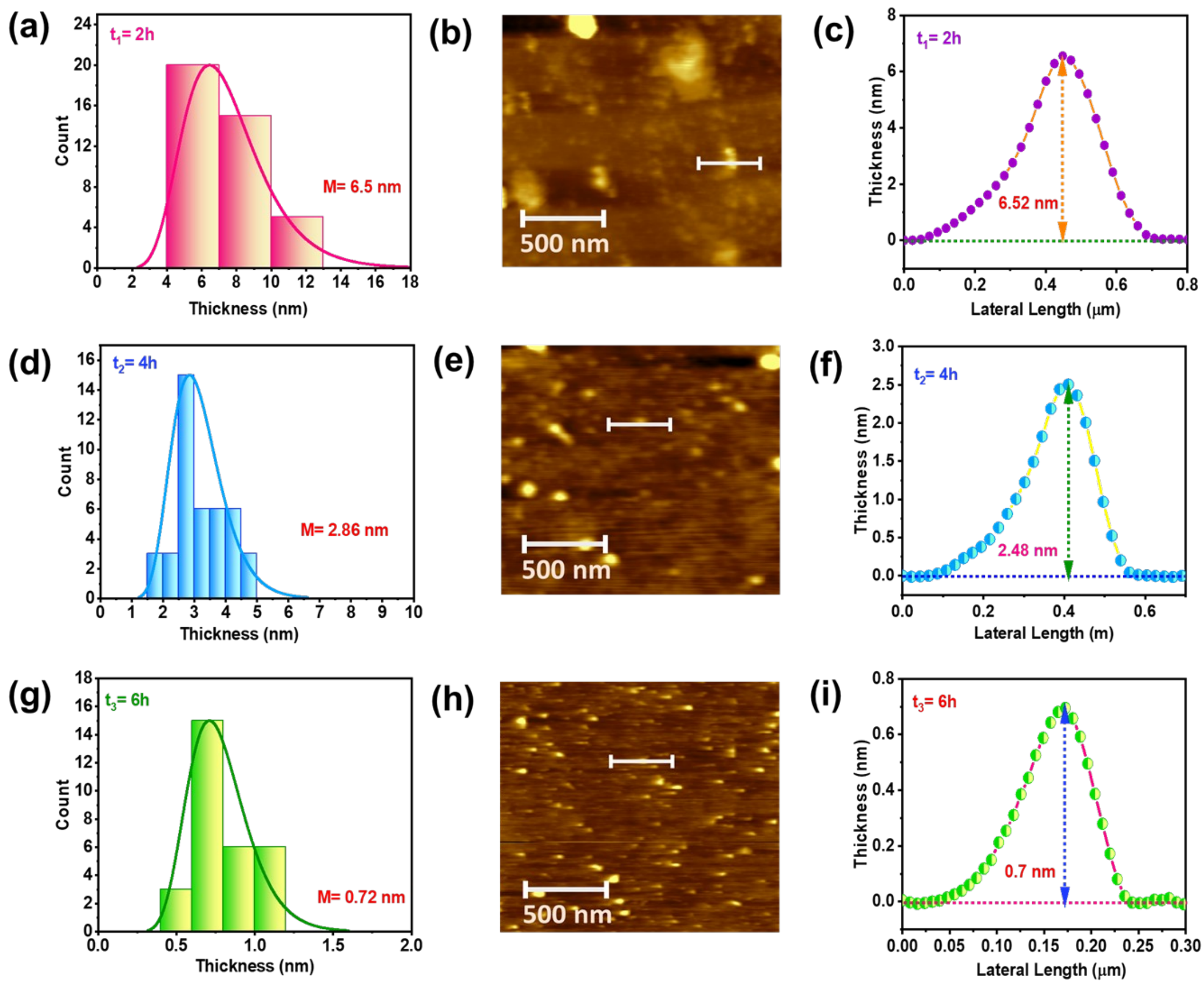

**Figure 1.** Thickness distribution characterization. Thickness distribution histograms of exfoliated muscovite nanoflakes are derived from AFM measurements, accompanied by representative images and height profiles at different exfoliation times: (a-c) for 2 h, (d-f) for 4 h, and (g-i) for 6h. M refers to the mean thickness value.

The degree of exfoliation is further analyzed using atomic force microscopy (AFM). Height profiles derived from AFM images are used to measure the lateral dimensions and thicknesses of samples exfoliated for 2, 4, and 6 hours. After 2 hours of exfoliation, the muscovite exhibits an average lateral dimension of approximately 0.93 µm **(Figure S2a**) and an average thickness of 6.5 nm, as shown in **Figure 1(a–c)**. With 4 hours of exfoliation, the average lateral size decreases to 0.56 µm **(Figure S2b)**, and the average thickness reduces to 2.86 nm **(Figure 1(d–f)).** Following 6 hours of exfoliation, the material shows a further reduction, with an average lateral dimension of 0.2 µm **(Figure S2c)** and an average thickness of 0.72 nm, indicating the formation of monolayers **(Figure 1(g–i))**.

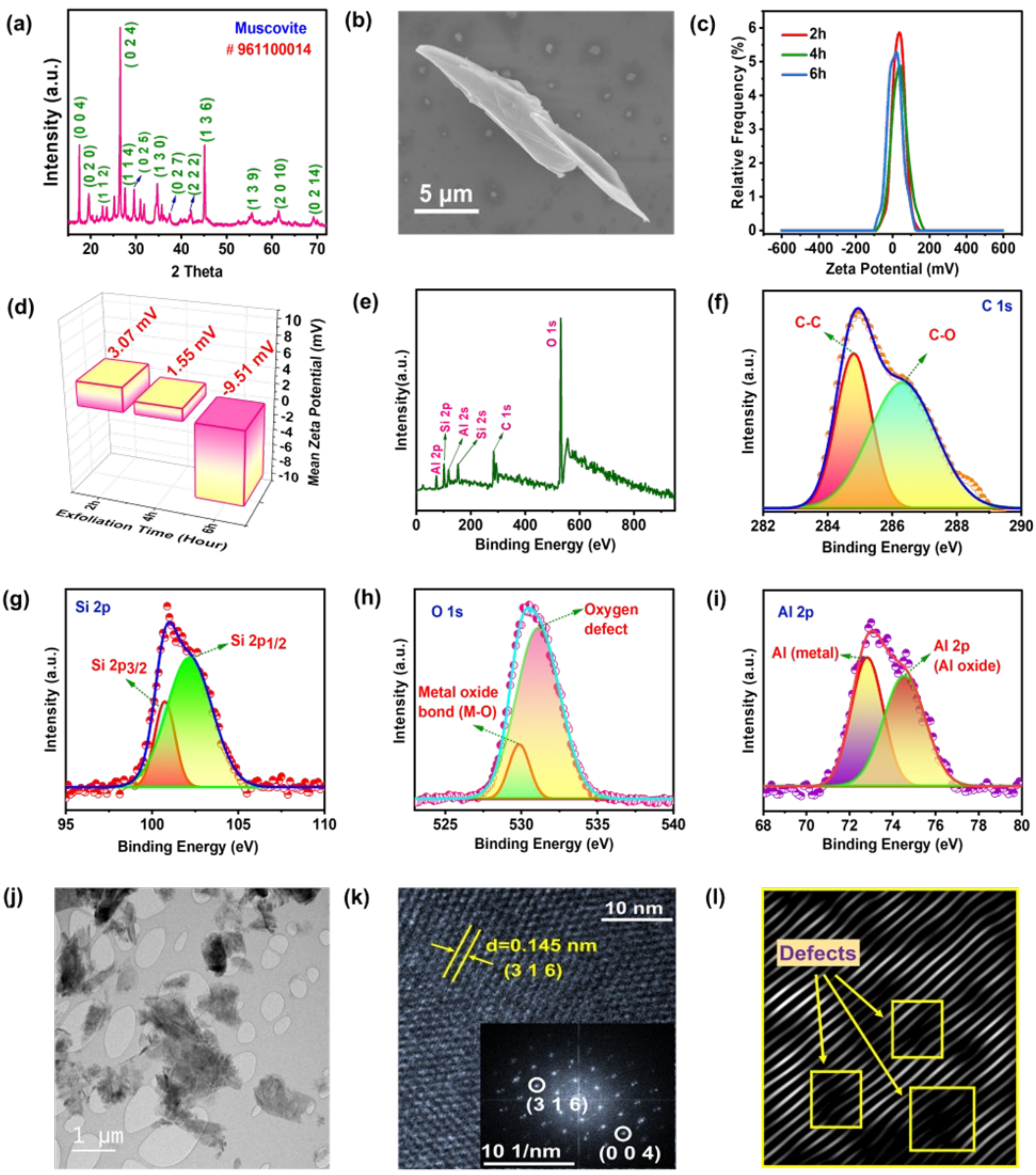

**Figure 2.** (a) XRD spectra of bulk muscovite; (b) SEM image of 2D muscovite; (c) Relative frequency vs. Zeta Potential plot for 2h, 4h, a nd 6h exfoliated muscovite (d) Bar plot of mean Zeta potential for 2, 4, and 6 h exfoliated muscovite; (e) The XPS surface scan of the 2D sample; (f-i) XPS spectra of individual peaks for C 1s, Si 2p, O 1s, Al 2p respectively; (j) Bright-field TEM image of 2D muscovite (k) The HRTEM image reveals various crystallographic plane orientations in 2D muscovite. The inset displays a Fast Fourier Transform (FFT) pattern, highlighting distinct spots that correspond to specific lattice planes present in the 2D muscovite structure; (l) The inverse FFT image, obtained by filtering a single spot from the FFT pattern, reveals lattice defects and dislocations in the 2D muscovite sample, highlighted by yellow-lined boxes.

The XRD pattern **(Figure 2a)** provides insights into the lattice planes and structural phases present in the bulk sample. Muscovite exhibits a monoclinic crystal structure, classified under the (C 1 2/c 1) space group. The primary crystallographic orientations correspond to the (0 0 4), (0 2 4), and (1 3 6) planes. The unit cell parameters are $a$ = 5.18 Å, $b$ = 9.02 Å, $c$ = 20.04 Å, with angles α = γ = 90° and β = 95.50°.

The elemental composition of the bulk muscovite is determined using an electron probe microanalyzer (EPMA). Based on the analysis, the calculated chemical formula for the bulk sample, as received, is

$$\left(K_{0.934}Na_{0.061}\right)_{\Sigma=0.994}\left(Al_{1.807}Fe_{0.109}Mg_{0.065}Ti_{0.024}Mn_{0.001}\right)_{\Sigma=2.006}\left(Si_{3.137}Al_{0.863}O_{10}\right)\left(OH\right)_2$$

To validate the computational approach, bulk muscovite is first examined **(Figure S3a)**. The calculated lattice parameters (a = 5.27 Å, b = 9.08 Å, c = 19.79 Å, α = γ = 90.00°, β = 95.95°) show excellent agreement with the experimental data, deviating by only 1.74%, 0.22%, and 0.95%, respectively. The results, summarized in **Table S1** in the supplementary information, also agree well with recent reported crystallographic data on muscovite [33–35]. Subsequently, a monolayer structural model is created by cleaving the bulk structure along the basal plane **(Figure S3b)**, allowing a clean separation with or without potassium passivation. Hydroxyl groups bonded to central Al atoms in the bulk are assumed to be removed during exfoliation. Retaining hydrogen atoms destabilizes the monolayer, as indicated by numerous imaginary vibrational modes. The optimized monolayer, consisting of 42 atoms, has lattice parameters: a = 5.24 Å, b = 9.01 Å, and a thickness of 7.76 Å, with angles α = γ = 90.00° and β = 97.26°. These values differ by less than 1.1% from the bulk, indicating structural integrity and that the isolated monolayer only slightly differs from the corresponding bulk layers. The computed thickness also aligns well with the experimental value of ~0.72 nm. Structurally, the monolayer is comprised of [$SiO_2$], [$SiO_3$], and [AlO] clusters, with average Si–O and Al–O bond lengths of ~1.60 Å and 1.70–1.95 Å, respectively. Ab initio molecular dynamics simulations are also performed to assess the thermodynamic stability of the muscovite monolayer. The minimal fluctuations in total energy during a 4.0 ps simulation at 800 K indicate that the structure remains stable under these conditions **(Figures S4a–S4c).**

**Figure 2(b)** shows an SEM image highlighting thin 2D muscovite flakes. The Energy Dispersive X-ray Spectroscopy (EDS) spectrum of the exfoliated sample, presented in **Figure S5,** reveals atomic percentages of 37.62% for C, 43.09% for O, 1.13% for Al, and 17.87% for Si. **Figure 2(c)** presents the distribution of relative frequency (%) with respect to the Zeta potential for 2D muscovite exfoliated at durations of 2, 4, and 6 hours. **Figure 2(d)** shows that the mean Zeta potentials for the 2h, 4h and 6h samples are 3.07 mV, 1.55 mV, and -9.51 mV, respectively. The progressive decrease in mean Zeta potential with increasing exfoliation time indicates a reduction in positive surface charge. This trend suggests the formation of potassium vacancies on the surface of the 2D muscovite flakes. **Figure 2(e–i)** present X-ray Photoelectron Spectroscopy (XPS) analyses of the 2D sample. The survey spectrum in **Figure 2(e)** identifies binding energy peaks corresponding to Al 2p (~74 eV), Si 2p (~101 eV), C 1s (~284.8 eV), Si 2s (~152 eV), and O 1s (~531 eV). **Figure 2(f),** the deconvoluted C 1s spectrum reveals a C–C bond at 284.8 eV and a C–O bond at 286.3 eV. The deconvoluted Si 2p spectrum in **Figure 2(g)** shows distinct peaks for Si $2p_{3/2}$ at 100.7 eV and Si $2p_{1/2}$ at 102.2 eV. **Figure 2(h)** displays the deconvoluted O1s spectrum, with a peak at 530.27 eV attributed to the metal oxide bond and another at 531.7 eV associated with oxygen defects[36]. **Figure 2(i)** presents the deconvoluted Al 2p spectrum, identifying peaks at 72.8 eV for metallic Al and 74.5 eV for Al in its oxidized state.

Following exfoliation, high-resolution transmission electron microscopy (HRTEM) is employed to examine the crystallographic orientation and surface defects of 2D muscovite. **Figure 2(j)** presents a bright-field TEM image, revealing thin, layered flakes. The corresponding dark-field HRTEM image in **Figure 2(k)** illustrates the atomic arrangement within the nanosheet. The inset of **Figure 2(k)** shows a Fast Fourier transform (FFT) of a selected region, displaying distinct diffraction spots. **Figure 2(l)** presents an inverse FFT image, generated by filtering a single diffraction spot, which highlights lattice imperfections and confirms the presence of surface defects in the 2D muscovite. The sonication process drives a structural transition from bulk muscovite to monolayer form. During liquid-phase exfoliation, shear forces separate the layers of the bulk material, producing few-layer muscovite. However, the high-energy input from prolonged sonication can displace atoms from their lattice positions, resulting in easily distinguishable surface defects.

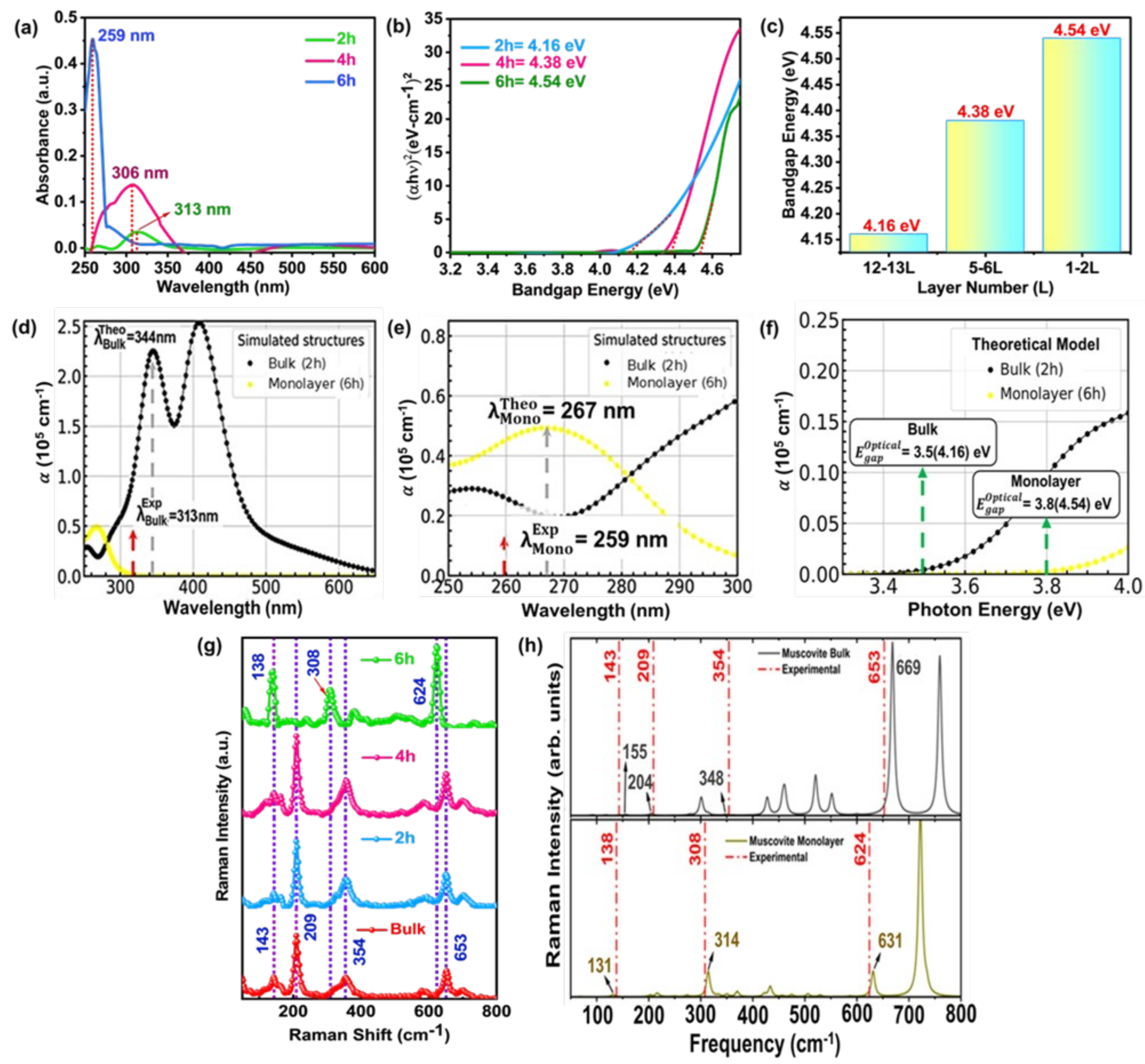

**Figure 3**. (a) Normalized linear absorption vs. exfoliation time; (b) The optical bandgap derived from experimental absorption data using the Tauc method for direct electronic transitions; (c) Bandgap energy as the function of layer numbers; (d) and (e) display the absorption spectra across wavelengths, enabling a direct comparison between experimental results ("Exp") and theoretical predictions ("Theo"); (f) presents the theoretical optical band gaps for both systems, with the experimentally obtained values provided in parentheses for reference; (g) Raman shift for bulk, 2h, 4h and 6h exfoliated muscovite; (h) Theoretical and experimental Raman spectra for comparison.

The ground-state absorption properties of 2D muscovite are investigated using ultraviolet–visible–near-infrared (UV–Vis–NIR) spectroscopy, as shown in **Figure 3(a)**. According to ligand field (or crystal field) theory, transition metal centers in suitable chemical environments absorb visible light through *d–d* electronic transitions, where 3d electrons are excited from the ground state to higher energy levels[37]. Exfoliated muscovite exhibits absorption peaks at 313 nm after 2 hours, 306 nm after 4 hours, and 259 nm after 6 hours of exfoliation. This progressive blue shift in the absorption band with increasing exfoliation time reflects a reduction in layer thickness and is attributed to quantum confinement effects[38]. The absorption spectra, plotted as a function of incident wavelength, reveal absorption onsets corresponding to optical band gap values of 4.16 eV, 4.38 eV, and 4.54 eV for direct electronic transitions, as illustrated in the Tauc plot in **Figure 3(b)**. **Figure 3(c)** presents the optical band gap energy as a function of layer numbers.

**Figure 3(d) and 3(e)** present the absorption coefficient (α) as a function of wavelength for the simulated (001) bulk and monolayer muscovite, using a spectral range comparable to that of the experimental data. For the bulk structure, the first prominent absorption peak deviates by approximately 11% from theoretical predictions, likely due to sample degradation occurring within the first two hours. In contrast, the monolayer simulation, focused on a narrower spectral range, captures the first absorption peak with a decreased error of around 3%, indicating an excellent agreement with the experimental observations. At longer wavelengths, the simulated spectra also closely match the experimental results.

To further validate the computational model, the optical band gap is estimated using density functional theory (DFT) simulations, resulting in values of 3.5 eV for bulk and 3.8 eV for monolayer muscovite (**Figure 3(f)**). These results are consistent with experimental trends. However, they are smaller than experimental reported values in the literature, which typically place the bulk muscovite band gap between 4.0 and 4.5 eV[39], reflecting a deviation larger than 10%. This band gap value underestimation is a well-known limitation of the generalized gradient approximation (GGA) using the PBE functional, which tends to underestimate band gap values[40]. As illustrated in **Figure 3(f)**, the discrepancy between experimental and DFT-derived optical band gaps arises primarily from this limitation. Nevertheless, the DFT results successfully capture the overall trend observed experimentally: the monolayer exhibits a larger band gap than the bulk, with both experimental and theoretical data showing an approximate 8% increase. This increase is attributed to quantum confinement, which increases the optical band gap as the material transitions from bulk to monolayer form.

**Figure S6 (a-c)** present the linear absorption coefficient (α), reflectivity (R), and refractive index (η) of the (001) monolayer and bulk muscovite as functions of photon energy in the range of 0 to 20 eV. The absorption coefficient for both bulk and monolayer structures remains nearly isotropic from the absorption onset up to approximately 11 eV. Although slight variations are observed among the three crystallographic directions, these differences are minimal and do not suggest any significant anisotropy in light absorption relative to polarization. Beyond 11 eV, corresponding to the deep ultraviolet region, distinct differences begin to appear. Around the first major absorption peak near 13 eV, the monolayer and bulk exhibit comparable absorption intensities. It is worth noting that in experimental contexts, absorption intensity can vary depending on sample preparation. However, for photon energies exceeding 13 eV, bulk muscovite demonstrates strong absorption. Notably, near 16 eV, the (001) monolayer shows a reduced absorption peak, while the bulk continues to exhibit increasing absorption. The refractive index (η) reaches its maximum around 5 eV for both structures and then gradually decreases with increasing photon energy. For the monolayer, the maximum refractive index decreases from 1.6 to 1.0, whereas for the bulk it decreases from 2.0 to 0.8. The maximum reflectivity (R) reaches approximately 24% for the bulk and 17% for the monolayer. These values, when considered alongside the refractive index data, suggest that a significant portion of incident light is absorbed in both the bulk and monolayer muscovite structures.

The stretching and vibrational modes of both bulk and exfoliated muscovite are analyzed using Raman spectroscopy, as shown in **Figure 3(g)**. The positions of all Raman peaks closely match those reported in previous studies [41]. Distinct spectral features are observed in the Raman spectra of both bulk and exfoliated muscovite:

(I) <600 $cm^{-1}$: The spectral features in this region stem from a multifaceted combination of translational motions of cations in octahedral and interlayer sites, along with interactions involving $SiO_4$ tetrahedra, $O_2$, and OH groups [41,42].

(II) 600-800 $cm^{-1}$: This spectral region is attributed to the vibrational modes of Si–$O_b$–Si bonds, where $O_b$ denotes bridging oxygen atoms that connect $SiO_4$ tetrahedra to form the layered framework. Muscovite exhibits a well-defined peak in the 650–700 $cm^{-1}$ range [41,42]. With increasing exfoliation time and decreasing flake size, from bulk to the 4-hour exfoliated sample, a noticeable increase in the Raman intensity is observed at 209 $cm^{-1}$, 354 $cm^{-1}$, and 653 $cm^{-1}$. This rise in intensity is associated with an increased level of structural disorder in the 2D muscovite [43]. The increased intensity is ascribed to variations in atomic bond lengths resulting from modified interlayer interactions. These structural modifications alter lattice vibrational dynamics, leading to the observed increase in Raman intensity.

In the sample exfoliated for 6 hours, a significant increase in Raman intensity is observed, along with a redshift of the 143 $cm^{-1}$ peak to 138 $cm^{-1}$. Additionally, the 209 $cm^{-1}$ peak disappears, while the 354 $cm^{-1}$ and 653 $cm^{-1}$ peaks shift to 308 $cm^{-1}$ and 624 $cm^{-1}$, respectively. These redshifts in the 354 $cm^{-1}$ and 653 $cm^{-1}$ modes are characteristic of the transition from bulk to monolayer muscovite.

The observed redshift in Raman spectra can be attributed to a combination of phonon confinement, surface relaxation, and lattice defects, all of which are increasingly significant at the nanoscale[44]. The total energy associated with lattice vibrations includes both interatomic binding energy and thermal vibrational energy. As particle size decreases, the surface-to-volume ratio increases, enhancing phonon confinement, a key factor influencing the physical properties of nanomaterials. This confinement raises the energy states of surface atoms, leading to larger atomic vibrational amplitudes and a corresponding reduction in vibrational frequency[44]. When the particle size drops below approximately 10 nm, comparable to the exciton diameter, these confinement effects become especially pronounced, resulting in a rapid decrease in Raman frequency. Such behavior indicates that Raman redshifts are driven by a combination of reduced dimensionality and surface-induced phonon relaxation. Furthermore, recent studies suggest that electron–phonon interactions become weaker as the size of nanoparticles decreases [45]. Nonetheless, structural defects introduced during synthesis play a pivotal role in altering the crystal lattice. These defects serve as efficient trapping sites for charge carriers, such as electrons, holes, and excitons, thereby significantly affecting both transport and optical properties[46]. The disruption of the ideal periodic lattice due to these imperfections leads to the formation of small nanocrystallites (NCs). In such NCs, the strict momentum conservation condition for Raman scattering ($q \approx 0$) is relaxed, allowing phonons away from the $\Gamma$ point to participate in the scattering process. This relaxation typically results in a redshift of Raman peaks, a phenomenon widely observed in nanostructured materials, such as nanowires, quantum dots, and defective graphene[46]. It is now well established that as the crystalline domain size decreases, Raman peaks in nanocrystalline materials exhibit a redshift in frequency[46]. This phenomenon likely contributes to the redshift observed in monolayer muscovite, where both size confinement and defect-related effects are at play.

The Raman spectra of bulk and monolayer muscovite are also simulated **(Figure 3(h))** and contrasted against the experimental findings, showing a good agreement with a maximum deviation of 8.4%.

An open-aperture (OA) Z-scan setup is employed to examine the layer-dependent nonlinear optical response of muscovite films in the femtosecond regime over a total Z-scan range of 120 mm. OA Z-scan measurements are conducted on muscovite samples exfoliated for 2, 4, and 6 hours and drop-cast onto glass substrates, using 450 nm excitation at varying laser intensities. A detailed overview of the Z-scan technique is provided in the experimental section. The following approximate equation is used to fit the normalized OA Z-scan data[12,47]

$$T(z) \approx 1 - \frac{\beta I_0 L_{eff}}{2^{\frac{3}{2}}\left(1+\frac{Z^2}{Z_R^2}\right)} \quad (1)$$

Here, β represents the TPA coefficient, and $L_{eff}$ denotes the effective sample length accounting for linear absorption. The imaginary component of the third-order nonlinear optical susceptibility is determined using the following expression [12]

$$\Im\left(\chi^{(3)}\right)(esu) = \frac{10^{-7} c\lambda n_0^2}{96\pi^2} \beta \quad (2)$$

Here, c, λ, and β are measured in units of cm $s^{-1}$, cm, and cm/W, respectively. To account for variations in linear absorption coefficients, a figure of merit (FOM) for third-order optical nonlinearity is introduced and defined as follows [12]

$$FOM\,(esu\,cm) = \frac{\Im\chi^{(3)}}{\alpha_0} \quad (3)$$

where $\alpha_0$ is the linear absorption coefficient.

**Figure 4(a)** presents representative OA Z-scan curves measured at different peak intensities for a muscovite film exfoliated for 2h, with an average thickness of 6.5 nm (approximately 12-13 layers), under 450 nm excitation. As verified in **Figure S8**, the substrate shows no evidence of nonlinear absorption. As the sample approaches the focal point, a noticeable drop in transmittance occurs, reaching its minimum at the focus, an indicative signature of reverse saturable absorption (RSA). The UV–Visible absorption spectrum of muscovite exfoliated for 2h shows a prominent peak at 313 nm, indicating a band gap of 4.16 eV. In contrast, the OA Z-scan experiment employs a 450 nm excitation beam (2.76 eV), where the photon energy is below the band gap but exceeds half its value ($\frac{1}{2}E_g < h\nu < E_g$) [47]. Under these conditions, the observed RSA is primarily attributed to two-photon absorption (TPA).

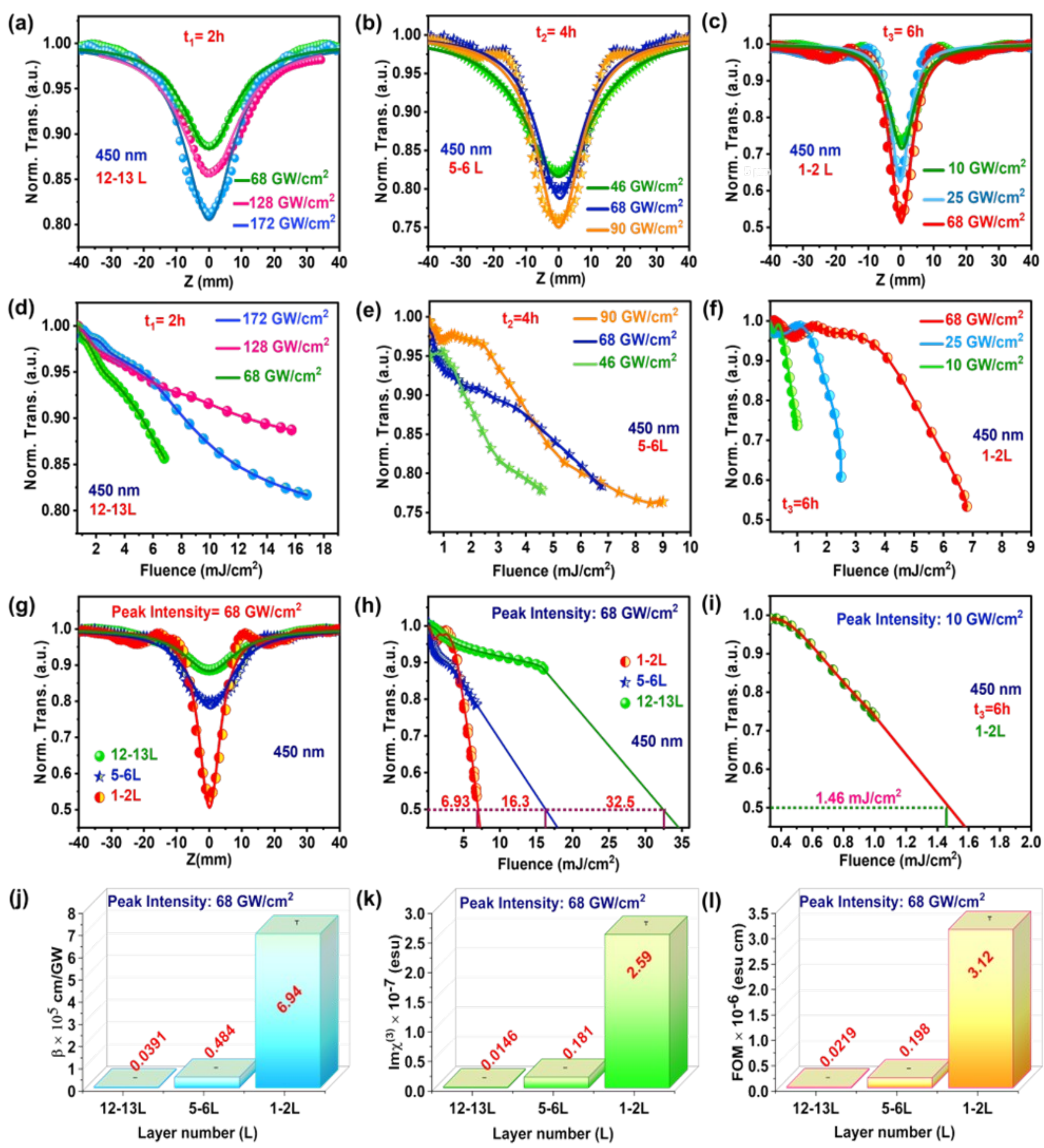

**Figure 4**. OA Z-Scan curves at 450 nm excitation wavelength of (a) 2h sonicated sample; (b) 4h sonicated sample; (c) 6h sonicated sample. OL curves at various peak intensities at 450nm excitation wavelength for (d) 2h sonicated sample; (e) 4h sonicated sample; (f) 6h sonicated sample. (g) Layer dependent TPA response at peak intensity 68 GW/cm$^2$ and 450 nm excitation wavelength. (h) Layer dependent OL threshold at peak intensity 68 GW/cm$^2$ and 450 nm excitation wavelength. (i) Lowest OL threshold achieved by monolayer muscovite at peak intensity 10 GW/cm$^2$ and 450 nm excitation wavelength. Comparison of (j) TPA coefficient (β); (k) the third order susceptibility Im ($\chi^{(3)}$); and (l) The figure of merit (FOM); as a function of layer numbers at common peak intensity 68 GW/cm$^2$ and 450 nm photoexcitation.

TPA is a nonlinear optical process in which two photons are absorbed simultaneously, exciting a molecule from a lower to a higher energy state. This can occur through a single optical field (degenerate TPA) or two distinct fields (nondegenerate TPA). The process is governed by symmetry selection rules that differ from those of one-photon absorption (OPA). These rules, formulated within the framework of the dipole approximation, have been thoroughly explored in previous studies [48–50]. According to these rules, in centrosymmetric systems, TPA transitions are allowed only between states of identical parity. For example, if the ground state possesses gerade (even) symmetry, TPA can only excite transitions to other gerade states, whereas ungerade (odd) states are accessible through one-photon absorption (OPA). In contrast, this parity restriction does not apply in non-centrosymmetric systems. However, the symmetry of the final electronic state, along with the polarization of the excitation beam(s), determines which components of the two-photon transition tensor are activated under specific experimental conditions [51]. Precise evaluation of a material's TPA characteristics is essential for its effective implementation in nonlinear optical applications. The TPA coefficient for the sample exfoliated for 2h is estimated to be $(9.25 \pm 0.15) \times 10^3$ cm/GW at a peak intensity of 172 GW/cm$^2$ and an excitation wavelength of 450 nm.

**Figure 4(b)** also illustrates the TPA behavior at 450 nm excitation wavelength for the muscovite sample exfoliated for 4h ($E_g$ = 4.38 eV), corresponding to approximately 5-6 layers. For this sample, the TPA coefficient $\beta$ is determined to be $(6.06 \pm 0.14) \times 10^4$ cm/GW at a peak intensity of 90 GW/cm$^2$. As shown in **Figure 4(c)**, the muscovite film exfoliated for 6h, corresponding to a monolayer, exhibits a further enhancement in nonlinear absorption. With an average thickness of 0.72 nm, the bandgap increased to 4.54 eV. OA Z-scan measurements at an excitation wavelength of 450 nm confirm the presence of TPA. Monolayer muscovite exhibits an exceptionally high TPA coefficient of $(6.94 \pm 0.17) \times 10^5$ cm/GW at a peak intensity of 68 GW/cm$^2$. This value surpasses those reported for widely studied 2D materials, such as graphene [52], $MoS_2$[21], and $WS_2$[22] by one to two orders of magnitude, and is comparable to the high performance observed in $PdSe_2$[53], as well as in 2D Biotite[12], a layered silicate structurally analogous to muscovite.

Based on our observations, the TPA response is triggered at a significantly lower intensity for the 6h exfoliated monolayer sample, starting at just 10 GW/cm$^2$, compared to the 2h exfoliated sample, where the response begins at around 68 GW/cm$^2$. Another key observation is that the 6h (1-2L) sample exhibits sharp and narrow TPA features, while the 2h (12-13L) and 4h (5–6L) samples display much broader TPA responses. This suggests that monolayer muscovite demonstrates a significantly stronger TPA behavior.

To investigate the layer-dependent TPA behavior, we select a common intensity of 68 GW/cm² and compare the TPA responses of the 2h (12-13L), 4h (5-6L), and 6h (1-2L) samples, as illustrated in **Figure 4(g)**. The marked increase in the TPA coefficient, from (3.91± 0.06) × $10^3$ cm/GW to (6.94 ± 0.17) × $10^5$ cm/GW, as the layer number decreases from 12-13L to 1-2L at peak intensity 68 GW/cm² and 450 nm excitation, highlights a strong layer-dependent nonlinear optical response, establishing monolayer muscovite as a highly promising candidate for ultrafast photonic and optoelectronic applications.

Prolonged exfoliation leads to a notable enhancement in nonlinear optical absorption, highlighting the critical influence of structural defects in tuning the material’s nonlinear optical behavior. These findings also demonstrate the promise of intrinsic muscovite for defect engineering, paving the way for its use in next-generation optoelectronic technologies.

**Table 1** summarizes the nonlinear optical parameters of 2D muscovite at various laser intensities under 450 nm excitation.

**Figure 4(d–f)** illustrate the intensity-dependent OL responses of muscovite samples exfoliated for 2, 4, and 6h respectively. The OL behavior of 2D muscovite is evaluated by fitting the normalized transmittance data to a polynomial function of position-dependent fluence. This approach facilitates the extraction of the OL threshold. The fluence as a function of position is computed using the following equation [12]:

$$F_{¿}(z)=\frac{4(\ln 2)^{1/2}E_{¿}}{\pi^{3/2}\omega_0^2\left(1+\frac{Z^2}{Z_R^2}\right)}\quad(4)$$

where $F_{¿}(z)$ is the position-dependent input fluence, and $E_{¿}$ is the input pulse energy.

Two key characteristics of an effective optical limiter are a low limiting threshold and a wide dynamic operating range. These are typically achieved through mechanisms such as reverse saturable absorption (RSA), two-photon and multiphoton absorption (TPA/MPA), FCA, NLR, and nonlinear scattering (NLS). In the case of 2D muscovite, the optical limiting response is predominantly attributed to TPA.

**Figure 4(h)** presents the layer-dependent OL thresholds for the 2h (12-13L), 4h (5-6L), and 6h (1-2L) samples at a common peak intensity of 68 GW/cm² and an excitation wavelength of 450 nm. The 2h exfoliated sample (12-13L) exhibits an OL threshold of 32.5 mJ/cm², which decreases to 16.3 mJ/cm² for the 4h exfoliated sample (5-6L), and further drops to 6.93 mJ/cm² for the 6h exfoliated sample (1-2L) under the same excitation conditions. Remarkably, monolayer muscovite achieves an even lower OL threshold of 1.46 mJ/cm² at a reduced peak intensity of 10 GW/cm² and 450 nm wavelength, as shown in **Figure 4(i)**. This enhanced OL behavior is directly correlated with the TPA coefficient (β), which quantifies the material's ability to simultaneously absorb two photons. A higher β value indicates stronger nonlinear absorption, enabling more efficient suppression of transmitted light under intense illumination. As exfoliation time increases from 2 to 6h, the muscovite flakes transition from multilayer to monolayer, accompanied by a substantial increase in β, from the $10^3$ to $10^5$ cm/GW range, resulting in a marked improvement in OL efficiency. Among all the samples, monolayer muscovite demonstrates the highest TPA coefficient and the lowest OL threshold, underscoring its exceptional nonlinear optical performance. This enhancement is attributed to two-photon resonance near the band edge, quantum confinement of charge carriers and defect-induced states introduced during the exfoliation process [54].

**Table 1.** Values of Nonlinear Optical Parameters for 2D Muscovite at various Laser Intensities at 450 nm, including β, $\Im(\chi^{(3)})$, FOM corresponding to 2h, 4h and 6h exfoliated Muscovite Nanoflakes.

| **Laser Parameters (Wavelengh, Pulse Width, Repetition Rate)** | **Exfoliation time & No. of Layers** | **Peak Intensity ($GW/cm^2$)** | **TPA Coefficient β (cm/GW)** | **Im $\chi^{(3)}$(esu)** | **FOM (esu cm)** | **Optical Limiting Threshold ($mJ/cm^2$)** |
|---|---|---|---|---|---|---|
| 450nm, 100fs, 1KHz | 2h 12-13 L | 68 | $(3.91 \pm 0.06)\times10^{3}$ | $(1.46\pm0.02)\times10^{-9}$ | $(2.19\pm0.03)\times10^{-8}$ | 32.5 |
| | | 128 | $(4.91 \pm 0.07)\times10^{3}$ | $(1.83\pm0.03)\times10^{-9}$ | $(2.75\pm0.04)\times10^{-8}$ | 33.05 |
| | | 172 | $(9.25 \pm 0.15)\times10^{3}$ | $(3.45\pm0.05)\times10^{-9}$ | $(5.18\pm0.08)\times10^{-8}$ | 34 |
| 450nm, 100fs, 1KHz | 4h 5-6 L | 46 | $(4.60 \pm 0.12)\times10^{4}$ | $(1.72\pm0.04)\times10^{-8}$ | $(1.88\pm0.04)\times10^{-7}$ | 14.03 |
| | | 68 | $(4.84 \pm 0.11)\times10^{4}$ | $(1.81\pm0.03)\times10^{-8}$ | $(1.98\pm0.04)\times10^{-7}$ | 16.3 |
| | | 90 | $(6.06 \pm 0.14)\times10^{4}$ | $(2.26\pm0.05)\times10^{-8}$ | $(2.48\pm0.06)\times10^{-7}$ | 17.5 |
| 450nm, 100fs, 1KHz | 6h 1-2L | 10 | $(2.68 \pm 0.07)\times10^{5}$ | $(1.01\pm0.02)\times10^{-7}$ | $(1.21\pm0.03)\times10^{-6}$ | 1.46 |
| | | 25 | $(3.16 \pm 0.07)\times10^{5}$ | $(1.18\pm0.03)\times10^{-7}$ | $(1.42\pm0.03)\times10^{-6}$ | 2.81 |
| | | 68 | $(6.94 \pm 0.17)\times10^{5}$ | $(2.59\pm0.06)\times10^{-7}$ | $(3.12\pm0.07)\times10^{-6}$ | 6.93 |

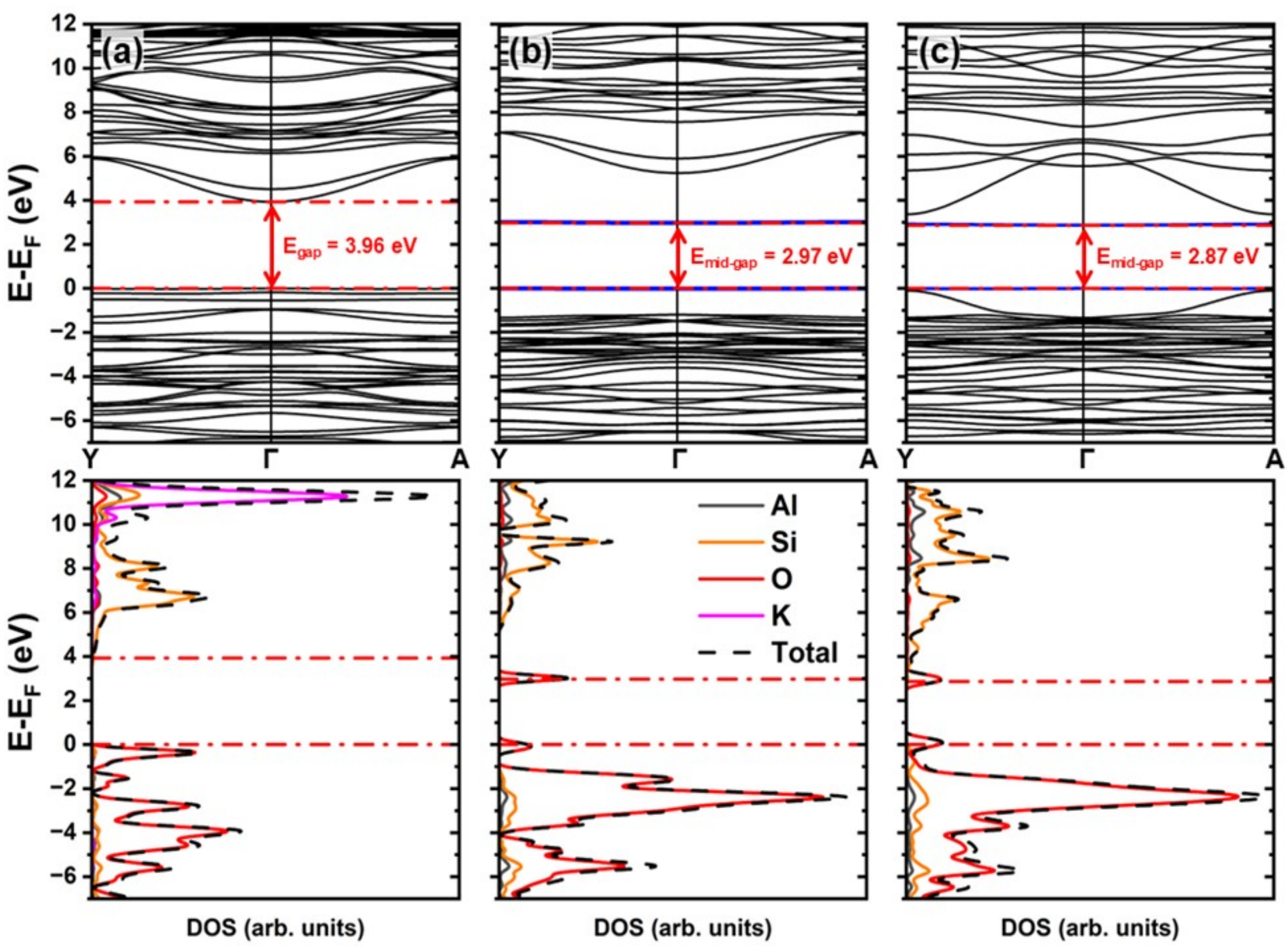

**Figure 5**. Electronic band structure (top) and the corresponding (bottom) Density of States (DOS) of (a) α-(001), (b) β-(001), and (c) γ-(001) muscovite monolayers.

Dopants and defects significantly influence the TPA process by introducing energy states within the bandgap or near the Fermi level. These mid-gap defect states act as intermediate levels, facilitating alternative electronic transitions and enhancing the TPA cross-section. To investigate the enhancement of the TPA process in the (001) muscovite monolayer, we have created three distinct configurations: (1) α-(001), a pristine monolayer with the potassium layer intact; (2) β-(001), where surface potassium atoms are removed; and (3) γ-(001), which features both potassium removal and an oxygen vacancy at the surface termination.
The transition from bulk muscovite to its monolayer form involves the removal of hydrogen atoms previously bonded to oxygen atoms, causing electronic distortions due to incomplete oxygen valencies. Additionally, prolonged sonication causes potassium ions ($K^+$) to leach from the muscovite surface, potentially creating oxygen vacancies and other defects. These vacancies form as oxygen atoms with unfulfilled valencies detach along with the liquid medium used in the sonication process. Consequently, defects and oxygen vacancies naturally arise during (001) muscovite sample preparation, creating mid-gap states that affect the TPA process.

Electronic structure analysis shows that removing the potassium layer introduces new states around the pristine band gap, generating mid-gap states. The band gap decreases from 3.96 eV in α-(001) to 2.97 eV in β-(001), and decreases further to 2.87 eV in γ-(001) when an oxygen vacancy is present. This decrease in the band gap value is due to the presence of localized mid-gap states, suggesting an enhancement of the TPA process, as it lowers the photon energy required for excitation; however, this band gap decrease alone does not fully explain the observed improvement. Another key observation is that the oxygen vacancy shifts the conduction band toward the mid-gap, making TPA more favourable by reducing the energy required after the first excitation.

The density of states (DOS) reveals that in α-(001) muscovite (**Figure 5a**), oxygen atoms dominate the valence band near the Fermi level, while silicon primarily contributes to the conduction band with small oxygen contributions. In β-(001) (**Figure 5b**) and γ-(001) (**Figure 5c**), mid-gap states appear, with oxygen significantly contributing to both the valence and conduction bands near the Fermi level, highlighting its critical role in the material's electronic behavior.

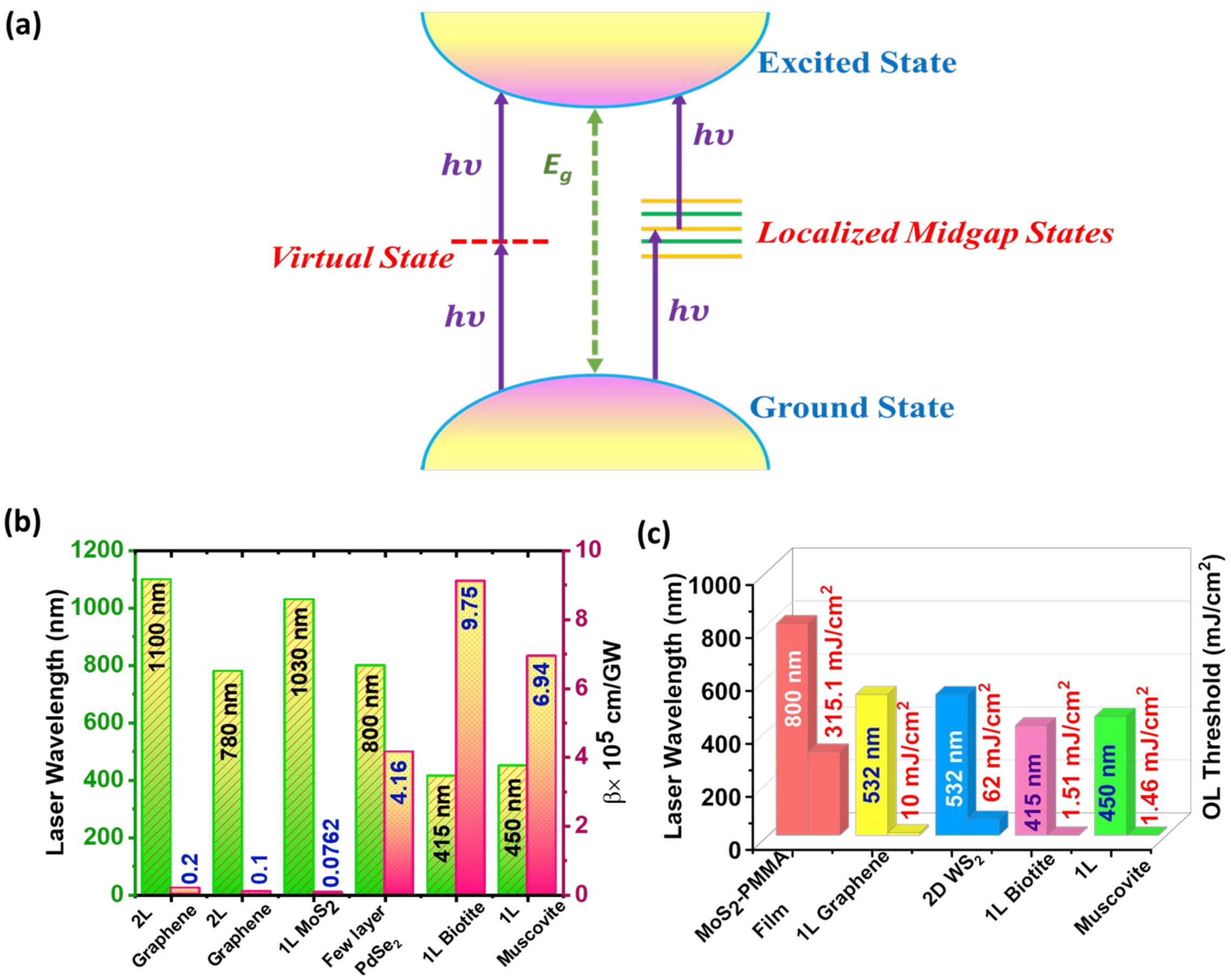

**Figure 6**. (a) Schematic of TPA process in 2D muscovite; (b) Comparison of the TPA coefficient (β) of other 2D materials with monolayer muscovite; (c) Comparison of OL threshold values for other 2D materials with monolayer muscovite.

**Figure 6a** illustrates the TPA process in 2D muscovite. DFT calculations confirm that TPA is significantly enhanced as the material transitions from a few-layer structure to a monolayer, due to the emergence of localized mid-gap states. These states are a result of extended liquid-phase exfoliation, which disrupts the potassium interlayer and creates oxygen vacancies. Raman spectroscopy, XPS, Zeta potential measurements, and HRTEM analysis further support the formation of such defects. These localized defect states act as intermediate energy levels, enabling two photons to simultaneously excite an electron from the ground state. Additionally, charge trapping at these defect sites increases the population of excited-state carriers, further enhancing the probability of TPA. This defect-assisted process leads to a pronounced nonlinear optical response[12].

**Figures 6(b) and 6(c)** showcase a comparative analysis of the TPA coefficient (β) and OL threshold of monolayer muscovite alongside other well-known 2D materials. Impressively, monolayer muscovite demonstrates a high TPA coefficient of 6.94 × $10^5$ cm/GW at 450 nm, closely matching that of $PdSe_2$ (4.16 × $10^5$ cm/GW at 800 nm) and monolayer Biotite (9.75 × $10^5$ cm/GW at 415 nm). Notably, it significantly outperforms bilayer graphene ((2 ± 0.4) × $10^4$ cm/GW at 1100 nm), monolayer $MoS_2$ ((7.62 ± 0.15) × $10^3$ cm/GW at 1030 nm) and monolayer $WS_2$ (1.183 × $10^3$ cm/GW at 800 nm). A comprehensive summary of these values is presented in **Table S2** of the supplementary information.

Monolayer muscovite exhibits an exceptionally low OL threshold of 1.46 mJ/cm² at 450 nm, outperforming several well-known 2D materials such as $MoS_2$-PMMA film (315.1 mJ/cm² at 800 nm), single-layer graphene (10 mJ/cm² at 532 nm), and $WS_2$ nanosheets (62 mJ/cm² at 532 nm). Its performance is also comparable to that of monolayer biotite (1.51 mJ/cm² at 415 nm). A detailed comparison of these values is available in **Table S3** of the supplementary information. The outstanding TPA coefficient combined with the superior OL response highlights the great promise of 2D muscovite for next-generation TPA-based photonic and optoelectronic applications.

## 3. Conclusions

In conclusion, this work demonstrates a novel strategy for synthesizing 2D muscovite directly from its natural ore using a controlled liquid-phase exfoliation process. AFM measurements confirm a systematic reduction in thickness from 6.5 nm to 0.72 nm, corresponding to a transition from approximately 12-13 layers to a monolayer. A gradual blueshift in the absorption band with decreasing layer number is observed, attributed to quantum confinement effects. Raman spectroscopy reveals significant structural evolution with exfoliation time. In the sample exfoliated for 6h, the 143 $cm^{-1}$ peak redshifts to 138 $cm^{-1}$ with an accompanying increase in intensity. As the material transitions from bulk to the 4h exfoliated state, peaks at 209 $cm^{-1}$, 354 $cm^{-1}$, and 653 $cm^{-1}$ show enhanced intensity. In the monolayer sample, the 209 $cm^{-1}$ peak disappears, while the 354 $cm^{-1}$ and 653 $cm^{-1}$ peaks shift to 308 $cm^{-1}$ and 624 $cm^{-1}$, respectively, signatures of the monolayer transition. These shifts are closely linked to the formation of surface defects during exfoliation, which disrupt the crystal lattice and produce nanocrystallites. In such structures, the relaxation of momentum conservation rules in Raman scattering ($q \approx 0$) allows phonons away from the Γ-point to contribute, leading to observable redshifts. The presence of structural defects is further corroborated by inverse FFT analysis of HRTEM images and the detection of oxygen vacancies in XPS spectra. The nonlinear optical properties of exfoliated muscovite are systematically investigated using the OA Z-scan technique. A remarkable enhancement in the TPA coefficient is observed as the layer number decreases from 12-13 layers to monolayer, rising from (3.91 ± 0.06) × $10^3$ cm/GW to an impressive (6.94 ± 0.17) × $10^5$ cm/GW under a common peak intensity of 68 GW/cm² at 450 nm excitation. These values significantly exceed those of widely studied 2D materials such as graphene, $MoS_2$, and $WS_2$. The OL threshold also exhibits strong layer dependence, decreasing from 32.5 mJ/cm² in 12-13 layer sample to just 6.93 mJ/cm² in monolayer sample at a peak intensity of 68 GW/cm$^2$ and 450 nm laser excitation. Notably, monolayer muscovite demonstrates an exceptionally low OL threshold of 1.46 mJ/cm² at 10 GW/cm$^2$ peak intensity and 450 nm laser excitation, outperforming many well-known 2D materials. These results highlight the outstanding potential of 2D muscovite for a broad range of advanced nonlinear optical applications, including two-photon microscopy (TPM), photodynamic therapy (PDT), optical data storage, optical limiting, optical switching, frequency conversion, and emerging fields such as quantum computing and information processing. This study opens new avenues for the development of naturally derived 2D silicates with superior nonlinear optical performance.

## 4. Experimental Section

**Synthesis Method:** A liquid-phase exfoliation technique is employed to derive 2D layers from bulk muscovite. In this process, 65 mg of bulk muscovite is dispersed in 60 mL of isopropyl alcohol (IPA) and sonicated at room temperature using a Rivotek probe sonicator for durations of 2, 4, and 6h. After sonication, the samples are left undisturbed for 24h to allow the precipitate to settle at the bottom of the container. The supernatant containing the exfoliated layers is then separated via centrifugation and used for further analysis.

**Characterization of Samples:** The XRD pattern of muscovite is recorded using a Bruker D8 Advance X-ray diffractometer equipped with a LynxEye detector. Cu Kα radiation with a wavelength of 0.15406 nm is used, and the measurements are carried out over a 2θ range of 20-100°. Absorption spectra are measured using an Analytical UV–Vis 2080Plus double-beam spectrophotometer with quartz cuvettes of 10 mm path length. Room-temperature Raman spectra are recorded using a WiTec UHTS 300 VIS Raman spectrometer (Germany), operated with an excitation wavelength of 532 nm. The sample composition is analyzed using a PHI 5000 VersaProbe III scanning X-ray photoelectron spectroscopy (XPS) microprobe. A Jeol JSM-IT300HR is used to capture SEM images showing thin flakes of 2D muscovite. An HRTEM FEI Themis 60−300, coupled with an FEI-CETA 4k × 4k camera, is used to examine crystallographic plane orientations and surface defects. The thickness of the 2D flakes is determined using an Oxford Asylum Research MFP 3D Atomic Force Microscope at room temperature.

**EPMA Analysis:** EPMA data acquisition was carried out at Rice University, Department of Earth, Environmental, and Planetary Sciences, using a JEOL JXA 8530F Hyperprobe equipped with a field-emission-assisted thermo-ionic (Schottky) emitter, five Wavelength-Dispersive Spectrometers (WDS), and one SD EDS detector.

**Z-scan and Laser Source:** The OA Z-scan technique is used to evaluate the nonlinear transmittance of the samples. A Ti:Sapphire femtosecond laser (Coherent Libra HE), operating at 800 nm with a pulse duration of 50 fs and a repetition rate of 1 kHz, serves as the main laser source. The output beam is directed into an optical parametric amplifier (OPA, TOPAS-Prime) to generate laser pulses with a broadly tunable central wavelength and a pulse duration of 100 fs for the Z-scan measurements.

The experimental setup, illustrated in **Figure S7**, involves splitting the input beam using a 90(T)/10(R) beam splitter. One portion is directed to a silicon photodetector (PD1), while the other is focused onto the sample using a plano-convex lens with a 20 cm focal length. The beam waist at the focal point and the corresponding Rayleigh range are determined. The sample, drop-cast onto a thin glass substrate (0.13 mm thick), is positioned at the focal point and translated along the z-axis using a motorized translational stage (Newport GST-150) controlled by a motion controller (Newport ESP-150).

To avoid photodetector saturation, variable neutral density filters are placed in front of the detectors. The transmitted beam is collected by a silicon photodetector (PD2, Thorlabs PDA100A-EC) with a 1.5 mm aperture placed in front to measure optical absorption. Variations in laser amplitude remain within 2%, exerting minimal impact on the Z-scan data. Signal quality is enhanced using 7225 DSP lock-in amplifiers, with an optical chopper placed before the sample to provide a reference frequency. Both the lock-in amplifiers and the translational stage are integrated with LabVIEW 2012 software for automated data acquisition.

## 5. Computational Methodology

The structural, electronic, and vibrational properties were investigated using periodic density functional theory within the CRYSTAL23 framework[55]. Calculations employed the PBE functional with the following basis sets: 86-311G** (Si)[56], 8-411d1 (O)[57], 86-511G (K),[58] 86-21G* (Al)[59], and 5-11G* (H)[60]. All stationary points were confirmed as minima via Hessian matrix diagonalization and by the absence of imaginary vibrational frequencies. Convergence criteria for the root mean square of the gradient components and nuclear displacements were set to 0.0001 and 0.0004 a.u., respectively. The accuracy of the Coulomb and Hartree-Fock exchange series was controlled by five integral tolerance parameters ($\alpha_i$, i = 1-5), set to 7, 7, 7, 7, and 16. A Monkhorst-Pack grid with a shrinking factor of 8 was used, and the electronic structure was determined along the high-symmetry k-point in the first Brillouin zone. Vibrational frequencies at the gamma point were calculated using numerical second derivatives of the total energies, estimated via the coupled perturbed Hartree-Fock/Kohn-Sham algorithm[61].

The optical properties of bulk and monolayer muscovite were investigated using ab initio simulations within the SIESTA code[62,63] framework. These simulations employed density functional theory[64] with the Perdew-Burke-Ernzerhof (PBE) exchange-correlation functional[65] under the generalized gradient approximation. A polarized double-zeta basis set of numerical orbitals was used, along with a mesh cutoff of 350 Ry. A Γ-centered Monkhorst-Pack grid[66] of 10x10x1 was employed for both bulk and monolayer systems. A 20 Å vacuum buffer was introduced along the z-direction for the monolayer simulations to prevent spurious interactions between mirror cells. Self-consistency was achieved when the difference between input and output density matrix elements was less than $10^{-4}$ and residual forces were below 0.01 eV/Å. To assess the structural thermal stability of the monolayer and the dynamics of surface potassium, *ab initio* molecular dynamics simulations were performed. These simulations employed a 4 ps NVT ensemble at 800K with 1 fs timesteps, using a Nosé-Hoover thermostat and a Γ-centered Monkhorst-Pack k-point grid.

To unveil the optical properties, we have calculated the absorption coefficient ($\alpha$), the reflectivity ($R$), and the refractive index ($n$), as functions of the photon energy frequency ($\omega$), using the following equations[62]:

$$\alpha(\omega)=\sqrt{2}\,\omega\left[\left(\epsilon_1^2(\omega)+\epsilon_2^2(\omega)\right)^{1/2}-\epsilon_1(\omega)\right]^{1/2} \quad (5)$$

$$R(\omega)=\left[\frac{\left(\epsilon_1(\omega)+i\epsilon_2(\omega)\right)^{1/2}-1}{\left(\epsilon_1(\omega)+i\epsilon_2(\omega)\right)^{1/2}+1}\right]^2 \quad (6)$$

$$n(\omega)=\frac{1}{\sqrt{2}}\left[\left(\epsilon_1^2(\omega)+\epsilon_2^2(\omega)\right)^{1/2}+\epsilon_1(\omega)\right]^2 \quad (7)$$

**Acknowledgements**

D.M. and P.K.D. thankfully acknowledge the UPM (SGDRI) project of IIT Kharagpur for all the equipments used in the Z-scan measurements. C.S.T. acknowledges DAE Young Scientist

Research Award (DAEYSRA) and AOARD grant no. FA2386- 21-1-4014 and Naval Research Board for funding support. C.S.T. acknowledges the funding support of AMT and Energy& Water Technologies of TMD Division of DST. Guilherme S. L. Fabris thanks the postdoc scholarship financed by the São Paulo Research Foundation (FAPESP) (process number 2024/03413-9) and R.B.O. thanks CNPq process numbers 151043/2024-8 and 200257/2025-0. Mateus M. Ferrer thank CNPq process number 406160/2023 and 314728/2023-0. M.L.P.J. acknowledges financial support from FAPDF (grant 00193-00001807/2023-16), CNPq (grant 444921/2024-9), and CAPES (grant 88887.005164/2024-00). Douglas S. Galvão acknowledges the Center for Computing in Engineering and Sciences at Unicamp for financial support through the FAPESP/CEPID Grant (process number 2013/08293-7). We thank the Coaraci Supercomputer for computer time (process number 2019/17874-0) and the Center for Computing in Engineering and Sciences at Unicamp (process number 2013/08293-4).

**Supporting Information**

Supporting Information is available from the author.